\documentclass[a4paper,12pt]{article}
\usepackage[utf8]{inputenc}
\usepackage{amsmath,amssymb,amsfonts,amsthm}

\usepackage{xcolor}
\usepackage[normalem]{ulem}
\usepackage{hyperref}

\renewcommand{\t}[1]{\tilde{#1}}

\usepackage{authblk}

\title{No-go theorem for static spherically symmetric configurations composed of two charged pressureless fluid species}

\author[1,2]{Andr\'es Ace\~na}
\author[1,2]{Bruno Cardin Guntsche}
\author[2]{Ivan Gentile de Austria}
\affil[1]{Instituto Interdisciplinario de Ciencias B\'asicas, CONICET, Mendoza, Argentina.}
\affil[2]{Facultad de Ciencias Exactas y Naturales, Universidad Nacional de Cuyo, Mendoza, Argentina.}

\date{}

\begin{document}

\maketitle

\begin{abstract}
    We present a no-go theorem for spherically symmetric configurations of two charged fluid species in equilibrium. The fluid species are assumed to be dusts, that is, perfect fluids without pressure, and the equilibrium can be attained for a single dust from the balance of electrostatic repulsion and gravitational attraction. We show that this is impossible for two dust species unless both of them are indistinguishable in terms of their electric charge density to matter density ratio. The result is obtained in the main three theories of mechanics, that is, in Newtonian Mechanics, in Special Relativity and in General Relativity. In particular, as charged dust solutions have been used to study the possibility of black hole mimickers, this result shows that such mimickers can not be constructed unless the underlying charged particle has the correct charge to mass ratio.
\end{abstract}


\section{Introduction}

The general physical setting that we study is matter interacting both gravitationally and electromagnetically. In particular, we want to consider equilibrium configurations. Within the Newtonian Mechanics (NM) formalism this means that the gravitational attraction needs to be exactly balanced by the electric repulsion. For the description of matter we use the approximation of dust, that is, a perfect fluid whose equation of state is simply that the pressure is zero, which comes from considering that the thermal energy is negligible. In order for the electromagnetic interaction to take place, the dust needs to be electrically charged. Therefore there are two densities to be known, the matter density, $\rho$, and the electric charge density, $\sigma$. In NM it is easy to see that if
\begin{equation}\label{relECD}
    \sigma=\pm\sqrt{4\pi\epsilon_0 G}\rho
\end{equation}
then any static distribution of matter is possible, as the gravitational attraction between any two elements of the fluid is balanced by the corresponding electric repulsion. In this case the fluid is referred to as electrically counterpoised dust (ECD). A bit surprising is the fact that the same happens in General Relativity (GR): any static distribution of ECD is a solution of the Einstein-Maxwell system of equations \cite{Das1962,Varela2003}. This has been exploited to test features of the theory, like constructing regular objects with diverging density \cite{BonnorWickramasuriya1972} or unbounded redshifts \cite{BonnorWickramasuriya1975}, to test the "hoop conjecture" \cite{Bonnor1998}, and to analyze the black hole limit for charged fluids \cite{Horvat2005,MeinelHutten2011}.

Although theoretically any static distribution of matter made of ECD is possible, the actual occurrence of such a distribution poses a strong difficulty, as there is no fundamental particle that has the correct relationship between electric charge and mass. Leaving aside the $\pm$ sign, which only amounts to the charge sign convention, equation \eqref{relECD} tells us that the relationship between charge density and matter density is fixed and has the particular value
\begin{equation}
    \frac{\sigma}{\rho} \approx 8.6\times 10^{-11}\, C \cdot kg^{-1}.
\end{equation}
This is tiny. For comparison, if we consider a gas made of protons we have
\begin{equation}
    \frac{e}{m_p} \approx 9.6\times 10^{7}\,C\cdot kg^{-1},
\end{equation}
or for single ionized lead atoms
\begin{equation}
    \frac{e}{m_{Pb}} \approx 4.7\times 10^{5}\,C\cdot kg^{-1}.
\end{equation}
Therefore, if we want to have ECD made of ionized hydrogen, we need to ionize only one in $10^{18}$ atoms, and for lead atoms the relationship is $2:10^{16}$. This responds to the fact that all known particles belong in two classes. In one class, the particles have no charge, and therefore the gravitational attraction can not be balanced by electric repulsion. In the other class, the electric repulsion is huge compared to the gravitational attraction. In summary, there is no naturally occurring fluid where \eqref{relECD} is satisfied.

If we want to pursue the previous line of thought, where we start with a neutral gas and ionize the right proportion of atoms in order for \eqref{relECD} to be satisfied, then instead of having one single species fluid we have two species. For the hydrogen example, one of the species would consist of neutral hydrogen with no charge, and the other would consist purely of ionized hydrogen, having a charge density much higher than required. To study the mechanical behaviour of this system it is not enough to consider only one species. It is not trivial to decide how can we physically tell if we are dealing with one or two species, or if separating one fluid into two components is unnecessary. It may seem that by making a thermodynamical average to arrive at a fluid description of matter then all the microscopic quantities can be averaged, and therefore we always end up with a description by a single species. This is only true if the interactions that we are considering can not distinguish those microscopic properties. For example, if only gravitational interaction is considered, then we can not separate the particles by their mass-charge ratio. If we also include electromagnetic interaction, as we do here, then said particles can be distinguished by the mass-charge ratio. In this case, the thermodynamical average can be made but only among the particles that share the same distinguishable properties. In conclusion, we are led to consider the case of two charged fluid species, and to see if it is possible to construct configurations where the required relationship \eqref{relECD} is satisfied "on average". If this is possible, then the argument of starting with a neutral fluid and ionizing the right amount of atoms has a solid base. If it is not possible, then there is no natural situation where ECD can be expected to occur.

Here we restrict our considerations to spherical symmetry. This problem, within GR and without spatial symmetries, has been solved in \cite{Acena2023}. Restricting to spherical symmetry permits dropping one technical assumption needed in GR. We consider three theories: NM, Nordstr\"om gravity in Special Relativity (SR) and GR. As we are dealing with electrically charged matter, we need to satisfy Maxwell equations in the setting of each theory.

The article is organized as follows. The considered problem and result are stated in Section \ref{secTheo}. Then, the result is proved using the formalism of NM in Section \ref{secNM}, of SR in Section \ref{secSR}, and of GR in Section \ref{secGR}. We conclude with a Discussion \ref{SecDisc}.

\section{Problem statement and no-go theorem}\label{secTheo}

We consider two electrically charged dust species, denoted $A$ and $B$, in a static and spherically symmetric distribution. This means that the density of matter and density of electric charge of the first species are described by the functions
\begin{equation}
    \rho_A(r),\quad \sigma_A(r),
\end{equation}
and for the second species
\begin{equation}
    \rho_B(r),\quad \sigma_B(r),
\end{equation}
where $r$ is the radial coordinate. The fluid species are subjected to gravitational and electromagnetic interaction.

\vspace{5pt}
\textit{\textbf{No-go theorem: }There is no static spherically symmetric distribution with two electrically charged dust species unless
\begin{equation}\label{eqTheo}
    \sigma_A=\pm\sqrt{4\pi\epsilon_0 G}\rho_A\,\mbox{ and }\,\sigma_B=\pm\sqrt{4\pi\epsilon_0 G}\rho_B.
\end{equation}}

There are a few remarks worth making. The first is that the same sign needs to be chosen in the equalities \eqref{eqTheo}, as both species need to have the same type of charge for gravitational and electrical forces to be balanced. Also, if equations \eqref{eqTheo} are satisfied, then the two species can not be distinguished by their mass-charge ratio, and for the setting that we are considering they are effectively only one species. That is why we name the result a no-go theorem, because if we can distinguish two species then there is no equilibrium. In terms of what the theorem affirms about the physical world, it states that it is not possible to construct star-like objects of ECD if there is no fundamental particle, atom or molecule with the correct mass-charge ratio.

\section{Newtonian Mechanics}\label{secNM}

In this section we consider the problem from the perspective of NM. Here and in Section \ref{secSR} we use SI units. The fundamental quantities to consider are the forces to which each fluid species is subjected, which in turn give us the accelerations of the fluid elements.  As we restrict the problem to spherical symmetry, we use spherical coordinates, and the functions that describe the dust species distributions are the corresponding matter and charge densities,
\begin{equation}
    \rho_A(r),\quad \sigma_A(r),\quad\rho_B(r),\quad \sigma_B(r).
\end{equation}
It is convenient to define two functions, the total mass and charge inside the radius $r$,
\begin{equation}\label{defM}
   M(r) := 4\pi\int_0^r \left(\rho_A(\t{r})+\rho_B(\t{r})\right)\t{r}^2d\t{r},
\end{equation}
\begin{equation}\label{defQ}
   Q(r) := 4\pi\int_0^r \left(\sigma_A(\t{r})+\sigma_B(\t{r})\right)\t{r}^2d\t{r}.
\end{equation}
If we consider an element of species $A$ situated at radius $r$ with volume $\delta V$, then the gravitational force is
\begin{equation}
    F_g = -G \frac{M(r)}{r^2}\rho_A(r)\delta V,
\end{equation}
while the electric force is
\begin{equation}
    F_e = \frac{1}{4\pi\epsilon_0} \frac{Q(r)}{r^2}\sigma_A(r)\delta V.
\end{equation}
In order for the fluid element to be in equilibrium we need $F_e=-F_g$, and therefore
\begin{equation}\label{eq1new}
    4\pi\epsilon_0 G M(r)\rho_A(r) = Q(r)\sigma_A(r).
\end{equation}
This is the equilibrium equation for the first fluid species. The same argument for the second species gives
\begin{equation}\label{eq2new}
    4\pi\epsilon_0 G M(r)\rho_B(r) = Q(r)\sigma_B(r).
\end{equation}
If we add \eqref{eq1new} and \eqref{eq2new} and use the definitions \eqref{defM} and \eqref{defQ} we get
\footnote{We denote by a prime the derivative with respect to $r$.}
\begin{equation}
    4\pi\epsilon_0 G M(r)M'(r) = Q(r)Q'(r),
\end{equation}
which , using that $M(0)=0$ and $Q(0)=0$, integrates to
\begin{equation}
    Q(r)=\pm\sqrt{4\pi\epsilon_0 G}M(r).
\end{equation}
If we insert this in \eqref{eq1new} and \eqref{eq2new} we obtain
\begin{equation}
    \sigma_A(r)=\pm\sqrt{4\pi\epsilon_0 G}\rho_A(r)\mbox{ and }\sigma_B(r)=\pm\sqrt{4\pi\epsilon_0 G}\rho_B(r).
\end{equation}
It is clear from \eqref{eq1new} and \eqref{eq2new} that in both equations the same sign needs to be chosen. The conclusion is that the relationship between charge density and mass density is fixed and the same for both fluid species, otherwise there is no equilibrium and the distribution is not static. This completes the proof in the Newtonian paradigm.

\section{Special Relativity}\label{secSR}

In SR we are concerned with the fields that describe the dynamical behaviour of the fluid species, that is, their four-velocities.  The other fields related to the species that we have to account for are the matter density and the electric charge density. So, we have
\begin{equation}\label{varSR}
    \rho_A,\quad\sigma_A,\quad u_A^\mu,\quad \rho_B,\quad\sigma_B,\quad u_B^\mu
\end{equation}
as the mass densities, charge densities and four-velocities of the fluid species. The electromagnetic theory was incorporated from the onset in SR, and in fact it was one of the main motivations for the development of SR. On the contrary, the incorporation of gravity into flat spacetime is not so straightforward. In this section we consider the problem in Nordstr\"om scalar theory of gravity. This theory is constructed in SR, the spacetime is Minkowski and we use standard Cartesian coordinates.
For a general discussion of Nordstr\"om theory and for the conventions in units and signs we follow \cite{Gourgoulhon2013}.
This theory of gravity allows the introduction of the gravitational interaction trough a scalar potential, whose source term is the trace of the energy-momentum tensor. It needs to be mentioned that Nordstr\"om theory is not a physically correct theory, as it is not compatible with observations. In particular, it predicts a periastron retardation instead of the observed periastron advance. Nevertheless, it is a useful middle step between NM and GR, which permits gaining insights within the relativistic setting.

For the electromagnetic field, we use the description through a vector potential, $A^\mu$, and gravity is given by a scalar potential, $\Phi$. The electromagnetic field is governed by Maxwell equations. The Faraday tensor is
\begin{equation}
    F_{\mu\nu}=\partial_\mu A_\nu - \partial_\nu A_\mu.
\end{equation}
For simplicity we use the Lorenz gauge, $\partial_\mu A^\mu=0$, and therefore the equation for $A^\mu$ is
\begin{equation}\label{eqA}
    \square A^\mu = -\frac{1}{\epsilon_0}j^\mu,
\end{equation}
where $\square$ is the d’Alembertian operator\footnote{In Minkowski spacetime the d’Alembertian operator is
\begin{equation}
    \square=-\frac{1}{c^2}\frac{\partial^2}{\partial t^2}+\frac{\partial^2}{\partial x^2}+\frac{\partial^2}{\partial y^2}+\frac{\partial^2}{\partial z^2}.
\end{equation}} and $j^\mu$ is the electric current density. With the notation \eqref{varSR} we have
\begin{equation}
    j^\mu = \sigma_A u_A^\mu + \sigma_B u_B^\mu.
\end{equation}

The governing equation for $\Phi$ is
\begin{equation}
    \square \Phi = -\frac{4\pi G}{c^2} T,
\end{equation}
where $T$ is the trace of the energy-momentum tensor.
The energy-momentum tensor is the sum of the energy-momentum tensor of the dust species and the energy-momentum tensor of the electromagnetic field. For dust, the energy momentum tensor is
\begin{equation}
    T_d^{\mu\nu} = T_A^{\mu\nu} + T_B^{\mu\nu} =  c^2\rho_A u_A^\mu u_A^\nu + c^2\rho_B u_B^\mu u_B^\nu.
\end{equation}
The electromagnetic energy-momentum tensor is 
\begin{equation}
    T_e^{\mu\nu} = \epsilon_0 \left(F_{\gamma}\,^\mu F^{\gamma\nu}-\frac{1}{4}F_{\gamma\lambda}F^{\gamma\lambda}g^{\mu\nu}\right).
\end{equation}
Taking the trace we have $T_d = -c^2(\rho_A + \rho_B)$ and $T_e=0$, therefore
\begin{equation}\label{eqPhi}
    \square \Phi = 4\pi G (\rho_A + \rho_B).
\end{equation}

As the last set of equations, we need the equations of motion for the fluid species. The Lorentz force in a fluid element is
\begin{equation}
    f_e^\mu = \sigma F^{\mu\nu}u_\nu,
\end{equation}
and the gravitational force, which comes from the scalar field interaction, is
\begin{equation}
    f_g^\mu = -\rho \partial_\nu \Phi (g^{\mu\nu}+u^\mu u^\nu)-\rho \Phi a^\mu,
\end{equation}
where $a^\mu$ is the four-acceleration of the fluid element. The equation of motion is
\begin{equation}
    \rho c^2 a^\mu = f_e^\mu + f_g^\mu.
\end{equation}
If we put this together for each fluid species, the equations are
\begin{equation}\label{eqUA}
    \rho_A (c^2 + \Phi) a_A^\mu = \sigma_A F^\mu\!_\nu u_A^\nu -\rho_A \partial_\nu \Phi (g^{\mu\nu}+u_A^\mu u_A^\nu),
\end{equation}
\begin{equation}\label{eqUB}
    \rho_B (c^2 + \Phi) a_B^\mu = \sigma_B F^\mu\!_\nu u_B^\nu -\rho_B \partial_\nu \Phi (g^{\mu\nu}+u_B^\mu u_B^\nu).
\end{equation}

Now that we have collected all the equations that have to be satisfied, we restrict them to the static case. In the first place this means that none of the functions depend on the time coordinate $t$. Also, the four-velocities and four-accelerations are
\begin{equation}
    u_A^\mu = u_B^\mu = \delta^\mu_0,\quad a_A^\mu = a_B^\mu = 0.
\end{equation}
Using these in \eqref{eqA} we have that\footnote{We denote by $\Delta$ the three-dimensional Laplacian, that is,
\begin{equation}
    \Delta = \frac{\partial^2}{\partial x^2}+\frac{\partial^2}{\partial y^2}+\frac{\partial^2}{\partial z^2}.
\end{equation}}
\begin{equation}
    \Delta A^\mu = -\frac{1}{\epsilon_0}(\sigma_A+\sigma_B)\delta^\mu_0
\end{equation}
and therefore
\begin{equation}
   A^\mu = V\delta^\mu_0,
\end{equation}
where $V$ is a function that does not depend on $t$ and satisfies
\begin{equation}\label{eqV}
    \Delta V = -\frac{1}{\epsilon_0}(\sigma_A+\sigma_B).
\end{equation}
With this $V$ the Lorenz gauge condition on $A^\mu$ is directly satisfied. Also, from \eqref{eqPhi},
\begin{equation}\label{eqPhiStatic}
    \Delta \Phi = 4\pi G (\rho_A + \rho_B).
\end{equation}
Completing the transition to the static case, \eqref{eqUA} and \eqref{eqUB} give\footnote{By $\nabla$ we denote the gradient:
\begin{equation}
    \nabla = \left(\frac{\partial}{\partial x},\frac{\partial}{\partial y},\frac{\partial}{\partial z}\right).
\end{equation}}
\begin{equation}\label{eqEqui}
    \rho_A \nabla\Phi = -\sigma_A\nabla V,\quad\rho_B \nabla\Phi = -\sigma_B\nabla V.
\end{equation}
It is interesting to note that the system of equations \eqref{eqV}-\eqref{eqPhiStatic}-\eqref{eqEqui} is the same as in the static case in NM.

Finally, if we also impose spherical symmetry, then all considered functions depend only on the radial coordinate, $r$, and
\begin{equation}
    \Delta \Phi = \frac{1}{r^2}(r^2\Phi')'.
\end{equation}
Then, \eqref{eqPhiStatic} can be integrated as
\begin{equation}
    \Phi' = \frac{4\pi G}{r^2}\int_0^r (\rho_A(\t{r})+\rho_B(\t{r}))\t{r}^2 d\t{r} = 4\pi G\frac{M(r)}{r^2}.
\end{equation}
Analogously, from \eqref{eqV},
\begin{equation}
    V' = -\frac{1}{\epsilon_0 r^2}\int_0^r \left(\sigma_A(\t{r})+\sigma_B(\t{r})\right)\t{r}^2d\t{r} = -\frac{1}{\epsilon_0}\frac{Q(r)}{r^2}.
\end{equation}
Inserting these last two equations in \eqref{eqEqui} we obtain
\begin{equation}
    4\pi\epsilon_0 G M \rho_A=Q \sigma_A,\quad 4\pi\epsilon_0 G M \rho_B=Q \sigma_B.
\end{equation}
These equations are the same as \eqref{eq1new}-\eqref{eq2new} and the result follows identically.

\section{General Relativity}\label{secGR}

In this section we attack the problem from the perspective of GR. We use geometrized units, where $G=c=1$, and also $\epsilon_0=(4\pi)^{-1}$. As in SR, for the description of the dust species the variables are the matter densities, charge densities and four-velocities, that is
\begin{equation}\label{varGR}
    \rho_A,\quad\sigma_A,\quad u_A^\mu,\quad\rho_B,\quad\sigma_B,\quad u_B^\mu.
\end{equation}
For the electromagnetic field, we use again the electromagnetic potential, $A_\mu$, and for the spacetime the fundamental object is the metric, $g_{\mu\nu}$.

The Einstein equations are
\begin{equation}
G_{\mu \nu} =  8\pi T_{\mu \nu},
\end{equation}
where $G_{\mu\nu}$ is the Einstein tensor,
\begin{equation}
   G_{\mu\nu}:=R_{\mu \nu } - \frac{1}{2} R g_{\mu \nu},
\end{equation}
being $R_{\mu\nu}$ the Ricci tensor and $R$ the curvature scalar. $T_{\mu\nu}$ is the energy momentum tensor, and in our case it has a contribution from the dust species and a contribution from the electromagnetic field,
\begin{equation}
    T_{\mu\nu} = T^d_{\mu\nu} + T^e_{\mu\nu}.
\end{equation}
The dust part is
\begin{equation}
    T^d_{\mu\nu} = T^A_{\mu\nu} + T^B_{\mu\nu} =  \rho_A u^A_\mu u^B_\nu + \rho_B u^B_\mu u^B_\nu.
\end{equation}
The electromagnetic energy-momentum tensor is 
\begin{equation}
    T^e_{\mu\nu} = \frac{1}{4\pi} \left(F_{\gamma\mu} F^{\gamma}\,_{\nu}-\frac{1}{4}F_{\gamma\lambda}F^{\gamma\lambda}g_{\mu\nu}\right),
\end{equation}
where the Faraday tensor is given by\footnote{We denote by $\nabla_{\mu}$ the torsion-free metric connection.}
\begin{equation}
    F_{\mu\nu}=\nabla_\mu A_\nu - \nabla_\nu A_\mu.
\end{equation}
Although not an independent equation, the energy-momentum conservation, $\nabla_\nu T^{\mu\nu} = 0$, is useful to simplify the calculations. This equation encodes an equation of motion if there is only one four-velocity in spacetime, meaning the presence of only one species. It constitutes the most direct and typical procedure for solving equilibrium equations for isolated systems in GR, and it can be a direct analogy to Newton's equations. However, in this case, it is merely a conservation equation, as happens in NM with energy conservation.

The Maxwell equations are
\begin{equation}\label{eqAgr}
    \nabla_\nu F^{\mu\nu} = 4\pi j^\mu,
\end{equation}
where $j^\mu$ is the current density. With the notation \eqref{varGR} we have
\begin{equation}
    j^\mu = \sigma_A u_A^\mu + \sigma_B u_B^\mu.
\end{equation}

As the last set of equations, we need the equations of motion for the fluid species, since the conservation equation does not give them individually. Although there is more than one way to arrive at said equations, the shortest path is to take the already known equation of movement for a fluid subjected to electromagnetic interaction in SR, and use the equivalence principle to generalize it to curved spacetime. The Lorentz force in a fluid element is
\begin{equation}
    f_e^\mu = \sigma F^{\mu\nu}u_\nu
\end{equation}
and the equation of motion is
\begin{equation}
    \rho u^\nu\nabla_\nu u^\mu = f_e^\mu.
\end{equation}
Then, for each fluid species
\begin{equation}
    \rho_A u_A^\nu\nabla_\nu u_A^\mu = \sigma_A F^{\mu}\,_\nu u_A^\nu,
\end{equation}
\begin{equation}
    \rho_B u_B^\nu\nabla_\nu u_B^\mu = \sigma_B F^{\mu}\,_\nu u_B^\nu.
\end{equation}

Now we impose staticity and spherical symmetry, considering that we use adapted coordinates, $(t,r,\theta,\phi)$, and therefore all the involved functions depend only on the radial coordinate $r$. In these coordinates, the metric has the form
\begin{equation}\label{metric}
   g_{\mu\nu}dx^\mu dx^\nu = -e^{2\Phi} dt^{2}+ e^{2 \Lambda} dr^{2} + r^{2} (d\theta^{2} + \sin^{2}\theta d\phi^{2}).
\end{equation}
As both species are static, then
\begin{equation}
    u_A^\mu = u_B^\mu = e^{-\Phi}\delta^\mu_0
\end{equation}
and
\begin{equation}
    j^\mu = (\sigma_A+\sigma_B)e^{-\Phi}\delta^\mu_0,
\end{equation}
which used in \eqref{eqAgr} shows that there is an "electrostatic potential", $V(r)$, such that
\begin{equation}
    A_\mu = V\delta^0_\mu, 
\end{equation}
and Maxwell equations simplify to
\begin{equation}\label{eqM}
      V'' + \left(\frac{2}{r} - \Lambda' -  \Phi'\right) V' =4\pi  e^{2 \, \Lambda + \Phi} (\sigma_A+\sigma_B).
\end{equation}

The non-zero Einstein equations are the $tt$, $rr$, $\theta\theta$ and $\phi\phi$ components, although the $\phi\phi$ is directly equivalent to $\theta\theta$. After some rearrangement, the Einstein equations are
\begin{equation}\label{eqE0}
     2r\Lambda' + e^{2\Lambda} - 1 - r^2e^{-2\Phi} V'^2 = 8\pi r^2 e^{2\Lambda} (\rho_A+\rho_B),
\end{equation}
\begin{equation}\label{eqE1}
    e^{2\Lambda} - 1 - 2r\Phi' = r^2 e^{-2\Phi} V'^2,
\end{equation}
\begin{equation}\label{eqE2}
    r\Phi''+ (1+r\Phi')(\Phi'-\Lambda') = r e^{-2\Phi} V'^2.
\end{equation}

The conservation equation is
\begin{equation}\label{eqC}
    V' V'' +  {\left(\frac{2}{r} - \Lambda' - \Phi' \right)} V'^2
    = 4\pi e^{2\Lambda + 2\Phi} \Phi'(\rho_A+\rho_B).
\end{equation}

For the fluid species, equations (56) and (57) become:
\begin{equation}\label{eqMov}
    \rho_A e^\Phi \Phi' = \sigma_A V',\quad
    \rho_B e^\Phi \Phi' = \sigma_B V'.
\end{equation}

If we multiply \eqref{eqE2} by $r$ and subtract \eqref{eqE1}, we get
\begin{equation}
    e^{2\Lambda}+r\Lambda'(1+r\Phi')=(1+r\Phi')^2+r(r\Phi''+\Phi'),
\end{equation}
which can be integrated as
\begin{equation}
    \Lambda=\ln(1+r\Phi').
\end{equation}
Substituting this in \eqref{eqE1} we get that
\begin{equation}\label{eqVgr}
    V = \pm e^\Phi.
\end{equation}
Finally, equations \eqref{eqMov} give
\begin{equation}\label{resultGR}
    \sigma_A = \pm\rho_A,\quad\sigma_B = \pm\rho_B,
\end{equation}
where the sign needs to be the same as in \eqref{eqVgr}. This completes the proof.

For completeness, if we multiply \eqref{eqM} by $V'$ and subtract \eqref{eqC}, then
\begin{equation}
    \sigma_A + \sigma_B = \pm(\rho_A+\rho_B),
\end{equation}
where again the sign needs to coincide with the sign in \eqref{eqVgr}, and is also a consequence of \eqref{resultGR}. To have the full set of equations, an equation for $\Phi$ can be obtained from \eqref{eqE0} or \eqref{eqC}. We get
\begin{equation}
    \rho_A + \rho_B = \frac{\Phi''+\Phi'(\Phi'+2/r)}{4\pi(1+r\Phi')^3}.
\end{equation}

\section{Discussion}\label{SecDisc}

We have presented and proved a no-go theorem in NM, SR and GR. The first remark to be made is that the extension of the theorem to more than two species is straightforward. The problem at hand, as it has the same stating and result for the three theories considered, highlights the particular perspective of each theory. In NM, the emphasis is on forces, and the equilibrium of forces is what is important for the equilibrium of the fluid elements. In SR, also forces are important, but the forces are the results of the fields, in this case the electromagnetic field and the scalar gravitational field. Finally, in GR there is no gravitational field, and therefore the fluid elements are accelerated, there is equilibrium because the electromagnetic force has the exact value as to produce the correct acceleration.

Our result implies that it is not possible to form spherically symmetric static ECD objects using charged dust species that do not satisfy \eqref{relECD} and averaging the mass and charge densities. Given that no known particle satisfies \eqref{relECD}, being the charge and mass not balanced by orders of magnitude, then star-like distributions of charged dust are not expected to occur.

In the context of GR, this means that black hole mimickers made of ECD are not viable. Also, extremal black holes are seen as the way of passing (or being an impassable barrier) between black holes and naked singularities. It seems that to form an extremal black hole by gravitational collapse it is necessary to start with a distribution of matter which is already extremal \cite{Wald1974}, \cite{Natario2016}, \cite{Sorce2017}. But it also seems that before this extremal matter limit is attained any object undergoes gravitational collapse, which strongly suggests that extremal Reissner-Nordstr\"om (ERN) black holes can not be produced by the collapse of charged spheres \cite{Anninos2001}. Then, ECD is the natural candidate to form an ERN black hole by collapse, being the relationship \eqref{relECD} the microscopic equivalent of the extremality condition $Q=M$. In this line, it has been shown that in the linear perturbations regime an ERN black hole can be formed from an ECD spacetime \cite{AcenaGentile2021}. The present result shows that unless there is a particle with the correct charge-mass ratio to start with this would not happen.

\section*{Acknowledgements}
The GR calculations were performed using SageMath \cite{SageMath} with the package SageManifolds \cite{SageManifolds}.


\end{document}